\documentclass[aps,pra,reprint,amsmath,amssymb,longbibliography]{revtex4-1}
\usepackage[T1]{fontenc}
\usepackage{lmodern}
\usepackage{graphicx,booktabs,array,bm}
\usepackage[colorlinks=true,linkcolor=blue,citecolor=blue,urlcolor=blue]{hyperref}
\hypersetup{pdftitle={Second-level termination of the PPT entanglement-cost hierarchy for two qutrits},pdfauthor={Wei Song}}
\newcommand{\Tr}{\operatorname{Tr}}
\newcommand{\diag}{\operatorname{diag}}
\newcommand{\ket}[1]{|#1\rangle}
\newcommand{\bra}[1]{\langle#1|}
\newcommand{\norm}[1]{\lVert#1\rVert}
\newcommand{\Cost}{E_{c,\mathrm{PPT}}^{\mathrm{exact}}}

\newtheorem{theorem}{Theorem}
\newtheorem{lemma}{Lemma}
\newtheorem{corollary}{Corollary}
\newenvironment{proof}[1][Proof]{\par\noindent\textit{#1.}\ }{\unskip\nobreak\hfill\quad$\square$\par}
\begin{document}
\title{Second-level termination of the PPT entanglement-cost hierarchy for two qutrits}
\author{Wei Song}
\email{wsong1@mail.ustc.edu.cn}
\affiliation{School of Physics and Materials Engineering, Hefei Normal University, Hefei 230601, China}
\author{Xiao-Lan Zong}\email{zxl@hfnu.edu.cn}
\affiliation{School of Physics and Materials Engineering, Hefei Normal University,
Hefei 230601, China}
\author{Ming Yang}
\affiliation{School of Physics, Anhui University, Hefei, 230601, China}
\affiliation{Leibniz International Joint Research Center of Materials Sciences of Anhui Province, Anhui University, Hefei, 230601, China}
\date{November 7, 2026}
\begin{abstract}
We prove that for arbitrary two-qutrit quantum states, the semidefinite programming lower-bound hierarchy for the PPT entanglement cost terminates at the second level. Therefore, under operations preserving positive partial transpose, its zero-error asymptotic entanglement cost can be obtained by a semidefinite program involving only three Hermitian matrix variables. The key to the proof is to shorten the feasible dual chain by one level; to this end, we combine finite certificates verified by exact rational arithmetic with analytic estimates near the boundary of the parameter domain. This result shows that a unified terminating level exists for two-qutrit systems, whereas no unified terminating level exists for all dimensions.
\end{abstract}
\maketitle

\section{Introduction}
We prove that for arbitrary two-qutrit quantum states, the semidefinite programming lower-bound hierarchy for the PPT entanglement cost terminates at the second level. Therefore, under operations preserving positive partial transpose, its zero-error asymptotic entanglement cost can be obtained by a semidefinite program involving only three Hermitian matrix variables. The key to the proof is to shorten the feasible dual chain by one level; to this end, we combine finite certificates verified by exact rational arithmetic with analytic estimates near the boundary of the parameter domain. This result shows that a unified terminating level exists for two-qutrit systems, whereas no unified terminating level exists for all dimensions.

Entanglement cost characterizes the asymptotic limit of the amount of entanglement resource consumed per copy on average when preparing multiple copies of a bipartite quantum state. For mixed states, this cost is not necessarily determined by the single-copy preparation problem. Under local operations and classical communication (LOCC), the entanglement cost with vanishing preparation error equals the regularized entanglement of formation~\cite{Hayden2001}. One can also allow operations that preserve positivity of the partial transpose (PPT)~\cite{Rains2001,Audenaert2003}. This paper studies the exact preparation problem under such operations. For each block length, the output state must exactly coincide with the corresponding tensor power of the target state, and the entanglement cost is defined as the asymptotic rate of resource consumption.

Wang and Wilde, in studying exact PPT preparation, introduced $\kappa$-entanglement computable by semidefinite programming (SDP)~\cite{Wang2020,Wang2023}. They later noted in an erratum that $\kappa$-entanglement is not additive, and in general the asymptotic entanglement cost equals its regularized value~\cite{Wang2024Errata}. Therefore, even if the single-copy quantity can be computed by SDP, it cannot directly determine the asymptotic entanglement cost.

Lami, Mele, and Regula~\cite{Lami2025} expressed this cost as the common limit of a hierarchy of SDP lower and upper bounds and gave convergence bounds, thereby providing an efficient method for approximating the entanglement cost. This raises the further question: is it sufficient to compute up to some fixed level to obtain the exact value of the cost? For $2\times d$ systems, their Corollary~1 shows that the first level of the lower-bound hierarchy already suffices. They also conjectured that for arbitrary bipartite quantum states, the second level of the lower-bound hierarchy gives the exact cost [Supplemental Material, Conjecture~S33].

Brinster et al.~\cite{Brinster2026} disproved this conjecture. They constructed quantum states for which adjacent levels take different values, and such examples can occur at arbitrarily high levels. The local dimensions of their analytic family of states increase with the level; in the example separating the second and third levels, the local dimensions on both sides are sixteen. However, these results still leave unanswered the question: when the local dimensions are fixed, does there exist a universal terminating level applicable to all quantum states?

This paper proves that for arbitrary two-qutrit quantum states, the lower-bound hierarchy terminates at the second level. Therefore, the entanglement cost of two-qutrit systems can be computed exactly by a fixed SDP, and there are no quantum states for which the second and third levels take different values. The key to the proof is to shorten an arbitrary feasible third-level dual chain by one level while keeping its first operator unchanged. This construction therefore applies to arbitrary input states. Equality of the values at the second and third levels in turn implies that all higher levels take the same value, yielding an exact expression for the entanglement cost.

The proof uses a computer-assisted approach. We first use spectral decomposition to reduce the construction problem to a vector with three Schmidt coefficients. Then we divide the parameter domain into two parts. In the region away from the boundary where the smallest Schmidt coefficient is zero, the required inequalities are guaranteed by finitely many positive-map certificates, all verified by exact rational arithmetic. Near the boundary, we give an analytic proof by jointly estimating the trace of a certain matrix block and several off-diagonal entries. Together, the two parts cover the entire parameter domain. The proof applies to both real and complex matrices and imposes no restriction on the rank of the quantum state.

\section{Hierarchy and main result}
\label{sec:main}
We use $\Gamma$ to denote partial transposition with respect to the second subsystem, and use $\norm{\cdot}$, $\norm{\cdot}_1$, and $\norm{\cdot}_F$ to denote the operator norm, trace norm, and Frobenius norm, respectively. Matrix inequalities are understood in the positive semidefinite order. The notation $A\succeq\pm B$ means that $A-B\succeq0$ and $A+B\succeq0$ hold simultaneously. All logarithms are base two.

PPT operations are completely positive trace-preserving maps $\Lambda$ such that $\Gamma\circ\Lambda\circ\Gamma$ is completely positive. If for all sufficiently large $n$ there exists a PPT operation $\Lambda_n$ such that
\begin{equation}
 \Lambda_n\bigl(\Phi_2^{\otimes\lfloor Rn\rfloor}\bigr)=\rho^{\otimes n},
\end{equation}
then $R$ is called an achievable rate for exact preparation of $\rho$. Here, $\Phi_2$ is the two-qubit maximally entangled state. We denote by $\Cost(\rho)$ the infimum of all achievable rates.

Following Definition S5 in Ref.~\cite{Lami2025}, let $S_{-1}=\rho$, and for $p\ge0$ define
\begin{equation}
 \chi_p(\rho)=\min\left\{\Tr S_p:
 S_i\succeq\pm S_{i-1}^\Gamma,\quad0\le i\le p\right\}.
 \label{eq:primal}
\end{equation}
The optimization variables are all Hermitian matrices, and the two inequalities at each step ensure that $S_i$ is positive semidefinite. For $q\ge1$, the upper-bound hierarchy is defined as
\begin{equation}
 \kappa_q(\rho)=\min\left\{\Tr S_{q-1}:
 \begin{array}{l}S_i\succeq\pm S_{i-1}^\Gamma,\quad0\le i<q,\\
 S_{q-1}^\Gamma\succeq0\end{array}\right\}.
 \label{eq:kappa}
\end{equation}
Let $E_{\chi,p}=\log_2\chi_p$ and $E_{\kappa,q}=\log_2\kappa_q$. The first level $E_{\kappa,1}$ of the upper-bound hierarchy is the $\kappa$-entanglement introduced in Refs.~\cite{Wang2020,Wang2023}. The entanglement cost satisfies the following upper and lower bounds~\cite{Lami2025}:
\begin{equation}
 E_{\chi,p}(\rho)\le\Cost(\rho)\le E_{\kappa,q}(\rho).
 \label{eq:sandwich}
\end{equation}

We write the dual problem as
\begin{equation}
 \chi_p(\rho)=\sup\left\{\Tr(\rho D_0^\Gamma):
 \begin{array}{l}I\succeq\pm D_p,\\
 D_{j+1}^\Gamma\succeq\pm D_j,\quad0\le j<p
 \end{array}\right\}.
 \label{eq:dual}
\end{equation}
This form can be obtained from Eq.~(S56) in Proposition~S15 of Ref.~\cite{Lami2025}: it suffices to set $D_i=V_i-W_i$, where $V_i,W_i$ are the positive semidefinite variables used there. Conversely, defining from $D_i$
\begin{align}
 V_i&=\tfrac12(D_{i+1}^\Gamma+D_i),&
 W_i&=\tfrac12(D_{i+1}^\Gamma-D_i),\quad i<p,\nonumber\\
 V_p&=\tfrac12(I+D_p),&W_p&=\tfrac12(I-D_p),
 \label{eq:dualconversion}
\end{align}
returns to the original form. The variable substitutions in both directions preserve feasibility and the objective value. From $\norm{\rho^\Gamma}\le\norm{\rho^\Gamma}_F\le1$, we know that $S_i=(i+2)I$ and $D_i=2^{i-p-1}I$ are strictly feasible solutions to the primal and dual problems, respectively. Therefore, the strict feasibility criterion for SDPs~\cite{Vandenberghe1996} guarantees strong duality, and this conclusion also applies to states that are not full rank.

The main result of this paper is given below.

\begin{theorem}[Hierarchy termination]
\label{thm:main}
For any quantum state $\rho$ on $\mathbb C^3\otimes\mathbb C^3$ and any $p\ge2$, we have
\begin{equation}
 \chi_p(\rho)=\chi_2(\rho).
 \label{eq:termination}
\end{equation}
Any feasible third-level dual chain can be shortened to a feasible second-level dual chain with the same $D_0$.
\end{theorem}

\begin{corollary}[Entanglement cost]
\label{cor:cost}
For any two-qutrit quantum state, we have
\begin{equation}
 \Cost(\rho)=E_{\chi,2}(\rho)=E_{\kappa,3}(\rho).
 \label{eq:cost}
\end{equation}
Therefore, the entanglement cost can be obtained by solving the following SDP and taking the base-two logarithm of its optimal value:
\begin{equation}
 \min\left\{\Tr S_2:
 S_0\succeq\pm\rho^\Gamma,\quad S_1\succeq\pm S_0^\Gamma,\quad
 S_2\succeq\pm S_1^\Gamma\right\}.
\end{equation}
This SDP contains only three $9\times9$ Hermitian matrix variables.
\end{corollary}

Below, we illustrate this result with two simple examples. For a PPT state, taking
$(S_0,S_1,S_2)=(\rho^\Gamma,\rho,\rho^\Gamma)$
yields a feasible chain with objective value one; hence $\chi_2=1$ and the entanglement cost is zero. For the maximally entangled state $\Phi_3=|\phi_3\rangle\langle\phi_3|$, where $|\phi_3\rangle=3^{-1/2}\sum_{j=0}^2|jj\rangle$, taking $S_0=S_1=S_2=I/3$ attains the lower bound $\chi_0(\Phi_3)=\norm{\Phi_3^\Gamma}_1=3$, so the entanglement cost is $\log_2 3$.

The following lemma shows that once adjacent levels take equal values, subsequent levels no longer change, and the entanglement cost can thus be determined. Proposition~5 and Eq.~(S142) of Ref.~\cite{Lami2025} used this criterion. For completeness, we give a proof according to the indexing conventions of this paper.

\begin{lemma}[Equality of adjacent levels]
\label{lem:plateau}
If for some $m\ge1$ we have $\chi_m(\rho)=\chi_{m-1}(\rho)$, then
\begin{equation}
 \chi_p(\rho)=\kappa_m(\rho)=\chi_{m-1}(\rho),\qquad p\ge m-1.
 \label{eq:plateau}
\end{equation}
In addition, $\Cost(\rho)=\log_2\chi_{m-1}(\rho)$.
\end{lemma}

\begin{proof}
The primal problem has an optimal solution. Indeed, the constraints give
$0\le\Tr S_0\le\cdots\le\Tr S_m$;
therefore, feasible solutions whose objective value is bounded by a given finite constant form a compact set. Take an optimal primal chain at level $m$; then
\begin{equation}
 \chi_{m-1}\le\Tr S_{m-1}\le\Tr S_m=\chi_m=\chi_{m-1}.
\end{equation}
Since $S_m-S_{m-1}^\Gamma$ is positive semidefinite and has trace zero, we must have
$S_m=S_{m-1}^\Gamma\succeq0$.
Therefore, after removing the last term $S_m$, the remaining chain is a feasible solution for $\kappa_m$, with objective value $\chi_{m-1}$.

Appending alternately $S_{m-1}^\Gamma$ and $S_{m-1}$ after this chain extends it to any higher level of the lower-bound hierarchy, while the objective value remains unchanged. Combining this with the monotonicity of $\chi_p$ and the additional constraint in $\kappa_m$ yields Eq.~\eqref{eq:plateau}. Finally, Eq.~\eqref{eq:sandwich} gives the expression for the entanglement cost.
\end{proof}

\begin{proof}[Proof of Corollary~\ref{cor:cost}]
Taking $m=3$ in Lemma~\ref{lem:plateau} and applying Theorem~\ref{thm:main} yields the conclusion.
\end{proof}

\section{Reduction of a dual chain}
\label{sec:shorten}
We shorten a feasible dual chain by one level while keeping \(D_0\) unchanged. In this way, regardless of the input state, the value of the objective function does not change. To this end, it suffices to construct a matrix \(X\) satisfying the conditions below.

Given a feasible dual chain \((D_0,D_1,D_2,D_3)\), let
\begin{equation}
 C=D_3^\Gamma-D_2,\qquad H=D_1^\Gamma,\qquad R=I-D_3.
 \label{eq:CHR}
\end{equation}
By the dual constraints,
\begin{equation}
 C,H\succeq0,\qquad H\succeq\pm D_0,\qquad I-(C+H)^\Gamma\succeq0.
 \label{eq:basic}
\end{equation}
The matrix \(X\) we seek must satisfy
\begin{align}
 &0\preceq X\preceq C,\qquad X\preceq2I,\label{eq:XY1}\\
 &P_X:=I+(C+H-2X)^\Gamma\succeq0.\label{eq:XY2}
\end{align}
Let \(Y=C+H-X\). From \(Y\succeq H\) and Eq.~\eqref{eq:basic}, we obtain \(Y\succeq\pm D_0\). Moreover,
\(I-X^\Gamma-Y^\Gamma=I-(C+H)^\Gamma\succeq0\).
These two relations, together with Eqs.~\eqref{eq:XY1} and~\eqref{eq:XY2}, guarantee that \((D_0,Y^\Gamma,I-X)\) is a feasible second-level dual chain.

If \(c:=\norm C\le2\), it suffices to take \(X=C\) directly. Then Eq.~\eqref{eq:XY1} clearly holds, and Eq.~\eqref{eq:XY2} follows from
\begin{equation}
 I-C^\Gamma+H^\Gamma=R+(D_2^\Gamma+D_1)\succeq0.
\end{equation}

It remains only to handle the case \(c>2\). Choose a normalized eigenvector \(v\) corresponding to the largest eigenvalue \(c\) of \(C\), and decompose \(C\) as
\begin{equation}
 C=C_0+E,\qquad C_0=c|v\rangle\langle v|=|r\rangle\langle r|.
 \label{eq:core}
\end{equation}
Thus \(E\succeq0\), \(C_0E=0\), and \(\norm E\le c\). In the Schmidt basis,
\begin{equation}
 r=x00+y11+t22,\qquad x\ge y\ge t\ge0,\quad c=x^2+y^2+t^2,
 \label{eq:r}
\end{equation}
where \(ij\) is shorthand for \(|ij\rangle\). Here, \(x,y,t\) are the Schmidt coefficients of the unnormalized vector \(r=\sqrt c\,v\). The boundary \(t=0\) mentioned below is the case in which the smallest Schmidt coefficient is zero.

This basis transformation also applies to complex matrices. Let \(L=U_A\otimes U_B\) and \(L_c=U_A\otimes\overline{U_B}\). Then
\begin{equation}
 (LML^\dagger)^\Gamma=L_cM^\Gamma L_c^\dagger.
\end{equation}
Applying the unitary conjugation by \(L\) to \(D_0,D_2\), and the unitary conjugation by \(L_c\) to \(D_1,D_3\), all dual constraints still hold, while \(C,H\) become \(LCL^\dagger,LHL^\dagger\), respectively. After constructing the shorter chain, one transforms back to the original basis, and its first operator is the original \(D_0\).

By Eq.~\eqref{eq:basic}, the expectation value of \(C+H\) on any normalized product vector is at most one. Since \(C_0\preceq C+H\), evaluating the expectation values on the Schmidt basis vectors and using \(c>2\), we obtain
\begin{equation}
 1\ge x\ge y\ge t>0,\qquad 2<c\le2+t^2\le3.
 \label{eq:domain}
\end{equation}
We seek \(X\) of the form
\begin{equation}
 X=sC_0+\alpha E,\qquad
 0\le\alpha\le s\le1,\qquad sc\le2.
 \label{eq:template}
\end{equation}
The supports of \(C_0\) and \(E\) are mutually orthogonal, so \(0\preceq X\preceq C\), and
\begin{equation}
 \norm X=\max\{sc,\alpha\norm E\}=sc\le2.
 \label{eq:templatenorm}
\end{equation}
Thus it remains only to prove that
\begin{equation}
 P_X=I+H^\Gamma+(1-2\alpha)E^\Gamma+(1-2s)C_0^\Gamma
 \label{eq:templateplus}
\end{equation}
is positive semidefinite.

When \(t\ge1/640\), the construction is completed by Lemma~\ref{lem:finite}.
When \(0<t\le1/640\), take \(s=1-t\), \(\alpha=0\), and apply Lemma~\ref{lem:boundary} to \(M=H+E\). In this case the norm constraint is also satisfied, because
\begin{equation}
 sc\le(1-t)(2+t^2)<2,
\end{equation}
where the last step uses
\(2-(1-t)(2+t^2)=t(2-t+t^2)>0\).

\begin{proof}[Proof of Theorem~\ref{thm:main}]
The direct construction for \(c\le2\) and the two lemmas for \(c>2\) cover all cases, and both lemmas include the endpoint \(t=1/640\). Therefore, every feasible third-level dual chain can be shortened to a second-level dual chain while keeping \(D_0\) unchanged. This shows that \(\chi_3\le\chi_2\), and monotonicity then gives \(\chi_3=\chi_2\). By Lemma~\ref{lem:plateau}, all higher levels have the same value.
\end{proof}

\section{Finite rational certificates}
\label{sec:finite}
This section proves the required inequality by means of finitely many certificates. Each certificate corresponds to a parameter region and writes the matrix in Eq.~\eqref{eq:XY2}, in a suitable basis, as a sum of several positive semidefinite terms and a remainder. We prove the positive semidefiniteness of the individual terms using rational Gram decompositions, and then use coefficient estimates to control the remainder over the entire region. Finally, we verify that these regions cover all feasible parameters away from the boundary.

\begin{lemma}
\label{lem:finite}
Let \((D_0,D_1,D_2,D_3)\) be a feasible chain of the third-level dual problem in Eq.~\eqref{eq:dual}, and adopt the spectral decomposition and Schmidt coordinates of Sec.~\ref{sec:shorten}. If \(c>2\) and \(t\ge1/640\), then one can choose parameters \(s,\alpha\) satisfying \(0\le\alpha\le s\le1\) and \(sc\le2\) such that \(X=sC_0+\alpha E\) satisfies Eqs.~\eqref{eq:XY1} and~\eqref{eq:XY2}.
\end{lemma}

\subsection{Positive inputs}
The dual constraints guarantee that the following operators are all positive semidefinite:
\begin{align}
 \mathsf Q&=I-C_0^\Gamma-E^\Gamma-H^\Gamma-R,\nonumber\\
 \mathsf B&=I-C_0^\Gamma-E^\Gamma+H^\Gamma-R,\nonumber\\
 \mathsf V&=2I-C_0-E-2R^\Gamma.
 \label{eq:positiveinputs}
\end{align}
In fact, they are respectively \(D_2^\Gamma-D_1\), \(D_2^\Gamma+D_1\), and \(D_3^\Gamma+D_2\). Likewise, \(H,E,R,2I-R\) are all positive semidefinite.

Take the following nine vectors as the column vectors of the matrix \(T\) in order:
\begin{equation}
 \begin{gathered}
 00,11,22,\quad01+10,02+20,12+21,\\
 01-10,02-20,12-21.
 \end{gathered}
 \label{eq:T}
\end{equation}
Define
\begin{align}
 N_T&=T^\dagger T=\diag(1,1,1,2,2,2,2,2,2),\nonumber\\
 Z_T(A)&=T^\dagger A^\Gamma T,\qquad n_i=(N_T)_{ii},\nonumber\\
 \lambda&=(x^2,y^2,t^2,xy,xt,yt,-xy,-xt,-yt),\nonumber\\
 C_*&=\diag(n_i\lambda_i),\qquad G_*=N_T-C_*,\nonumber\\
 J_*&=2\diag(1-xy,1-xt,1-yt).
 \label{eq:certificatebasis}
\end{align}
The last three basis vectors span the antisymmetric subspace, and we denote the matrix block corresponding to this subspace by the subscript \(--\). Below we will also need the inequality
\begin{equation}
             J_*+Z_{T,--}(H+E)\succeq0.
             \label{eq:skewpositive}
\end{equation}
To this end, let \(M_*=C+H\); then \(M_*\succeq0\) and \(M_*^\Gamma\preceq I\). Take any normalized vector of Schmidt rank two, which can be written as \(s00+u11\) in its local Schmidt basis, where \(s,u\ge0\) and \(s^2+u^2=1\). From the positive semidefiniteness of \(M_*\) and the upper bounds on its diagonal entries in the product basis, we obtain
\begin{equation}
 \bra{s00+u11}M_*^\Gamma\ket{s00+u11}
 \ge-2su|(M_*)_{01,10}|\ge-1.
\end{equation}
Since every \(3\times3\) complex antisymmetric matrix must be singular, the Schmidt rank of any vector in the antisymmetric subspace is at most two. Therefore, the above estimate guarantees that the compression of \(I+M_*^\Gamma\) onto this subspace is positive semidefinite, and Eq.~\eqref{eq:skewpositive} follows immediately. The condition that the local dimension is three is used here.

Each certificate uses eleven completely positive maps, whose output dimension is nine. The maps and their inputs are denoted by
\begin{equation}
\begin{array}{c|c@{\qquad}c|c}
 A_H&H&A_{HT}&H^{\mathsf T}\\
 A_E&E&A_{ET}&E^{\mathsf T}\\
 A_R&R&A_{RT}&R^{\mathsf T}\\
 U&2I-R&Q&T^\dagger\mathsf Q T\\
 B&T^\dagger\mathsf B T&V&\mathsf V\\
 S&J_*+Z_{T,--}(H+E)&&
\end{array}
\label{eq:maps}
\end{equation}
where the input dimension of \(S\) is three, and the input dimensions of the remaining maps are all nine. The full transpose preserves positive semidefiniteness, so the transposed inputs in the table remain positive semidefinite also in the complex-matrix case. Each Choi matrix is a direct sum of several rational Gram matrices \(LL^{\mathsf T}\). Under the index convention \((p,b)\leftrightarrow9p+b\), the corresponding maps are defined by
\begin{equation}
 \Phi(A)_{bd}=\sum_{p,q}A_{pq}J_{(p,b),(q,d)}.
 \label{eq:choi}
\end{equation}
The positive semidefiniteness of the Choi matrix implies that this map is completely positive~\cite{Choi1975}. This can also be seen directly from the Gram decomposition: rearranging each column of the Gram factor into a matrix \(K\) according to \(K_{bp}=L_{(p,b)}\), the map term corresponding to this column is \(KAK^\dagger\).

\subsection{Matrix residual and scalar certificates}
In Eq.~\eqref{eq:templateplus}, set \(\beta=2s-1\). The certificate file stores \(\alpha,\beta\), while \(s\) is determined by \(s=(1+\beta)/2\). Below, three quantities are used to control the error: \(\delta\) for matrix coefficient identities, \(\eta\) for the row sums of the absolute values of off-diagonal entries, and \(\varepsilon_{\rm pol}\) for scalar polynomial identities. These error bounds hold throughout the entire parameter region corresponding to each certificate.

Summing the outputs of the maps in Eq.~\eqref{eq:maps}, denote the result by \(\mathcal P\). Since all inputs are positive semidefinite, \(\mathcal P\succeq0\). The part that does not contain the matrix variables \(H,E,R\) is
\begin{equation}
 \begin{split}
 \mathcal P_0={}&Q(G_*)+B(G_*)+2U(I)\\
 &+V(2I-C_0)+S(J_*).
 \end{split}
\end{equation}
Including the remaining terms, we have
\begin{equation}
 \mathcal P=\mathcal P_0+Z_T(H)+(1-2\alpha)Z_T(E)+\mathcal E.
 \label{eq:defect}
\end{equation}
Rational maps do not necessarily satisfy the required coefficient identities exactly, so the error term \(\mathcal E\) cannot be discarded and must be estimated explicitly. For each matrix variable, the coefficient error corresponds to a complex linear map \(\mathcal R\) that preserves Hermiticity. In particular,
\begin{equation}
 \mathcal R(E_{ji})=\mathcal R(E_{ij})^\dagger.
 \label{eq:paired}
\end{equation}
Therefore, for any Hermitian input \(F\),
\begin{equation}
 \begin{split}
 \mathcal R(F)={}&\sum_i F_{ii}\mathcal R(E_{ii})\\
 &+\sum_{i<j}\bigl[F_{ij}\mathcal R(E_{ij})
 +\overline{F_{ij}}\mathcal R(E_{ij})^\dagger\bigr].
 \end{split}
\end{equation}
Thus, it suffices to know the values of \(\mathcal R\) on the \(45\) matrix units with \(i\le j\) in order to determine the entire error map. However, when estimating the error, both members of each adjoint pair must be included simultaneously in order to cover all complex off-diagonal entries. From the nonnegativity of the diagonal entries of \(\mathsf Q\) and the positive semidefiniteness of the inputs, it follows that \(|H_{ij}|,|E_{ij}|,|R_{ij}|\le1\). Summing the absolute values of the coefficient errors in each row then gives the rational upper bound \(\norm{\mathcal E}\le\delta\).

Let \(\mathcal D=N_T-\beta C_*-\mathcal P_0\). Substituting the above expression yields the exact identity
\begin{equation}
 T^\dagger(I-X^\Gamma+Y^\Gamma)T
            =\mathcal P+\mathcal D-\mathcal E.
 \label{eq:residualidentity}
\end{equation}
Every matrix element of \(\mathcal D\) is a linear combination of the following monomials:
\begin{equation}
                     1,x^2,y^2,t^2,xy,xt,yt.
\end{equation}
Denote the corresponding coefficient matrices by \(A_k\), and let
\begin{equation}
 \eta=\max_i\frac1{n_i}\sum_{j\ne i}\sum_k|(A_k)_{ij}|.
 \label{eq:eta}
\end{equation}
Then, throughout \([0,1]^3\), the sum of the absolute values of the off-diagonal entries in each row, divided by \(n_i\), does not exceed \(\eta\).

Consider the rectangle \([l,u]\times[a,b]\) in the \((x,t)\) plane, and let \(L_0=\max(a,1-b^2)\). All feasible parameters in this rectangle satisfy
\begin{equation}
 \begin{gathered}
 l\le x\le u,\quad L_0\le y\le u,\quad a\le t\le b,\\
 x\ge y\ge t,\qquad \sigma=x^2+y^2+t^2-2\ge0.
 \end{gathered}
 \label{eq:physicalbox}
\end{equation}
Here, the lower bound on \(y\) follows from \(y^2\ge1-t^2\). Let \(z_0=(1,x,y,t)^{\mathsf T}\), and define eight linear polynomials that are nonnegative on this feasible region:
\begin{equation}
 g=(x-l,u-x,y-L_0,u-y,t-a,b-t,x-y,y-t).
\end{equation}
For \(p_i=\mathcal D_{ii}/n_i\), the certificate gives
\begin{equation}
 \begin{split}
 p_i={}&\mu+z_0^{\mathsf T}G_i z_0+\kappa_i\sigma+w_{i0}
       +\sum_j w_{ij}g_j\\
      &+\sum_{j\le k}w_{ijk}g_jg_k+e_i(x,y,t).
 \end{split}
 \label{eq:scalarcertificate}
\end{equation}
The certificate data include a rational Gram decomposition of \(G_i\), and all weights and \(\kappa_i\) in the formula are nonnegative. Using the ten monomials of degree at most two as a basis, we compute the coefficients of the error polynomial \(e_i\) exactly. For each \(e_i\), we sum the absolute values of its coefficients and take the maximum of these sums, denoting it by \(\varepsilon_{\rm pol}\). Since all parameters lie between zero and one, we have \(p_i\ge\mu-\varepsilon_{\rm pol}\) throughout the entire region.

By the diagonal dominance criterion for symmetric matrices,
\(\mathcal D\succeq(\mu-\varepsilon_{\rm pol}-\eta)N_T\).
Combining \(N_T\succeq I\) and \(\norm{\mathcal E}\le\delta\), Eq.~\eqref{eq:residualidentity} gives
\begin{equation}
 I-X^\Gamma+Y^\Gamma\succeq
       (\mu-\varepsilon_{\rm pol}-\eta-\delta)I.
 \label{eq:certificatemargin}
\end{equation}
This estimate follows directly from exact coefficient bounds, and therefore holds at every point in the region.

\subsection{Coverage and exact checks}
We divide the parameter domain into three parts. For \(1/10\le t\le13/16\), we perform a binary partition of the rectangle
\begin{equation}
              (x,t)\in[13/16,1]\times[1/10,13/16],
\end{equation}
with all partition endpoints rational. Within this range of \(t\), all feasible parameters lie in this rectangle; otherwise \(x<13/16\), which would imply \(c\le3(13/16)^2<2\), contradicting \(c>2\). The partition tree has \(243\) nodes in total, of which \(119\) leaf nodes are verified by certificates, and another three leaf nodes are excluded. These three leaf nodes all satisfy \(2u^2+b^2<2\) and cannot contain feasible parameters. Each partition chooses one coordinate direction and splits the region into two halves at the exact midpoint of the corresponding interval. The verification program checks the two subregions and all endpoints obtained from each partition, and verifies all leaf nodes and certificates to ensure that nothing is omitted.

On each verified leaf node, take \(\alpha=1/2\), \(\beta=1-(a+b)/2\), and let \(s=(1+\beta)/2\). From the certificate data one can verify that
\begin{equation}
 (1+\beta)(2u^2+b^2)\le4,\qquad
 \mu-\varepsilon_{\rm pol}-\eta-\delta\ge\frac1{5000}.
\end{equation}
Combining this with \(c\le2u^2+b^2\), the first inequality gives \(sc\le2\). The chosen parameters also directly satisfy the remaining conditions in Eq.~\eqref{eq:template}.

For \(1/640\le t\le1/10\), we use the following six successively halved intervals:
\begin{equation}
 T_k=\frac1{10\,2^k},\qquad t\in[T_k/2,T_k],\quad0\le k\le5.
\end{equation}
Since \(x^2>1-t^2/2\), the feasible parameters corresponding to each interval all lie in the rectangle
\begin{equation}
 (x,t)\in[1-T_k^2/2,1]\times[T_k/2,T_k].
\end{equation}
Take \(\alpha=1/2\), \(\beta=1-3T_k/4\); each certificate then gives the lower bound \(1/100000\) in Eq.~\eqref{eq:certificatemargin}. At the same time,
\begin{equation}
 sc\le\tfrac12(2-3T_k/4)(2+T_k^2)\le2,
\end{equation}
so the norm constraint is also satisfied. These six closed intervals are consecutive and cover the entire \([1/640,1/10]\) without gaps.

Finally, for \(t\ge13/16\), take \(\alpha=s=2/3\), i.e., \(\beta=1/3\), so that \(X=2C/3\). From \(c\le3\) we immediately obtain \(sc\le2\). The entire cube \([13/16,1]^3\) can be covered by a single certificate. In this region, instead of Eq.~\eqref{eq:scalarcertificate}, we give a lower bound for each normalized diagonal entry by estimating the polynomial coefficients term by term. The coefficients of the resulting lower-bound polynomials are all rational, and all coefficients of nonconstant terms are nonnegative. Therefore, they all attain their minimum at the vertex \((13/16,13/16,13/16)\), where the smallest value is \(50331/2560000\). After subtracting the upper bounds for the off-diagonal terms and coefficient errors, the remaining lower bound is still greater than \(1/100\).

\begin{table}[t]
\caption{Parameter ranges covered by the certificates and the corresponding lower bounds for \(I-X^\Gamma+Y^\Gamma\). Each lower bound holds throughout the corresponding region.}
\label{tab:certificates}
\begin{ruledtabular}
\begin{tabular}{lcc}
Range of \(t\) & Number of certificate groups & Lower bound\\
$[1/640,1/10]$ & 6 & $I/100000$\\
$[1/10,13/16]$ & 119 & $I/5000$\\
$[13/16,1]$ & 1 & $I/100$
\end{tabular}
\end{ruledtabular}
\end{table}

All \(126\) certificate groups have been verified using exact rational arithmetic. Combining the above lower bounds, the region coverage check, and the norm estimates proves Lemma~\ref{lem:finite}. The verification procedure is described in Appendix~\ref{app:verification}, and the data, programs, and complete run logs are available in Ref.~\cite{Supplement}.

\section{The boundary region}
\label{sec:boundary}
We now consider the case in which the smallest Schmidt coefficient is close to zero. In this case, the required inequality can be proved directly by analytic methods, without resorting to finite certificates.

\begin{lemma}[Positive definiteness near the boundary]
\label{lem:boundary}
Let \(0<t\le\tau:=1/640\), and let \(x,y\in[0,1]\) satisfy \(x^2+y^2+t^2>2\).
Let \(r=x00+y11+t22\) and \(C_0=|r\rangle\langle r|\).
If
\begin{equation}
 M\succeq0,\qquad Q_0=I-C_0^\Gamma-M^\Gamma\succeq0,
 \label{eq:boundaryhyp}
\end{equation}
then
\begin{equation}
 P=I+M^\Gamma-(1-2t)C_0^\Gamma\succ0.
 \label{eq:boundarytarget}
\end{equation}
\end{lemma}

\subsection{Blocks and a joint trace estimate}
Let \(\mathcal O=\operatorname{span}\{00,01,10,11\}\), and define
\begin{align}
 \mathcal S&=\operatorname{span}\{00,11,(01+10)/\sqrt2\},\nonumber\\
 a&=(01-10)/\sqrt2,\qquad f=22,\nonumber\\
 \mathcal L_A&=\operatorname{span}\{02,12\},\quad
 \mathcal L_B=\operatorname{span}\{20,21\},\nonumber\\
 \mathcal L&=\mathcal L_A\oplus\mathcal L_B.
\end{align}
Each subspace uses the vectors listed above as an orthonormal basis, in the order given. Clearly,
\(\mathcal O=\mathcal S\oplus\operatorname{span}\{a\}\).
Below, subscripts denote the corresponding matrix blocks or compression operators in these bases. Let \(U_0=M_{\mathcal O\mathcal O}\) and \(\Theta=\Tr U_0\). Write the compression of \(M\) on \(\mathcal L\oplus\operatorname{span}\{f\}\) as
\begin{equation}
 M_{(\mathcal L,f)}=\begin{pmatrix}A&Z&h_A\\Z^\dagger&B&h_B\\h_A^\dagger&h_B^\dagger&p\end{pmatrix},
 \quad K_0=\begin{pmatrix}A&Z\\Z^\dagger&B\end{pmatrix}.
\end{equation}
Also define
\begin{align}
 N_{ij}&=M_{ij,22},\quad i,j\in\{0,1\},\nonumber\\
 h'&=(h_A,\overline{h_B})^{\mathsf T},\quad z=(Z_{01}-Z_{10})/2,\nonumber\\
 q&=(Z_{00},Z_{11},(Z_{01}+Z_{10})/\sqrt2)^{\mathsf T}.
\end{align}
After computing the partial transpose, the corresponding matrix blocks of \(P\) are
\begin{align}
 P_{\mathcal L\mathcal L}&=\begin{pmatrix}I+A&L\\L^\dagger&I+B^{\mathsf T}\end{pmatrix},\nonumber\\
 L&=N-(1-2t)t\diag(x,y),\nonumber\\
 P_{\mathcal L f}&=h',\quad P_{af}=\sqrt2z,\quad P_{\mathcal S f}=q,\nonumber\\
 P_{ff}&=1-(1-2t)t^2+p.
 \label{eq:PTblocks}
\end{align}

Let \(I_{\mathcal O}\) be the identity operator on \(\mathcal O\), and let it be zero on the orthogonal complement of \(\mathcal O\). Take \(\Omega=00+11\); then the projector onto the symmetric subspace \(\mathcal S\) satisfies
\begin{equation}
 \Pi_{\mathcal S}^\Gamma=\tfrac12(I_{\mathcal O}+|\Omega\rangle\langle\Omega|).
\end{equation}
Since the compression of \(Q_0\) on \(\mathcal S\oplus\operatorname{span}\{f\}\) is positive semidefinite, and \(0\le(Q_0)_{ff}\le1\), we have
\begin{align}
 \norm q^2&\le\Tr(Q_0)_{\mathcal S}\nonumber\\
 &\le(1-x^2)+(1-y^2)+(1-xy)-\Theta/2.
\end{align}
Here the second step uses
\(\Tr(U_0^\Gamma)_{\mathcal S}=(\Theta+\langle\Omega|U_0|\Omega\rangle)/2\ge\Theta/2\).
On the other hand, \(xy\ge\sqrt{1-t^2}\) gives
\begin{equation}
 1-xy\le\frac{t^2}{2-t}\le(1/2+t)t^2.
\end{equation}
Combining these estimates, we obtain
\begin{equation}
 \Theta+2\norm q^2\le(3+2t)t^2,\qquad0\le\Theta<4t^2.
 \label{eq:jointbudget}
\end{equation}
Next, examining the compressions of \(Q_0\) on \(\mathcal L_A\) and \(\mathcal L_B\) gives \(0\preceq A,B\preceq I\); combined with \(M\succeq0\), this yields \(\norm Z\le1\). In addition, we have \(0\le p\le1\), \(\norm N\le\sqrt{p\Theta}\le2t\), and \(\norm L\le3t\).

\subsection{An estimate for the antisymmetric component}
Let \(\zeta=|z|\le1\) and \(w=1-\zeta^2\). From the positive semidefiniteness of the compression of \(Q_0\) on \(\operatorname{span}\{a,f\}\), together with \((Q_0)_{aa}\le2+\Theta\le2+4t^2\), we obtain
\begin{equation}
 p\le w+t^2.
 \label{eq:pbound}
\end{equation}
Below we need the following trace inequality for block matrices:
\begin{equation}
 \begin{pmatrix}A&Z\\Z^\dagger&B\end{pmatrix}\succeq0
 \quad\Longrightarrow\quad
 2\norm Z_1\le\Tr A+\Tr B.
 \label{eq:blocktrace}
\end{equation}
This inequality can be proved directly. Take the left and right singular vector pairs \((u_j,v_j)\) of \(Z\), evaluate the quadratic form of the above positive semidefinite matrix on the vector \((u_j,-v_j)\), and then sum over all singular vector pairs. On the other hand, the duality relation for the trace norm gives
\(2\zeta=|Z_{01}-Z_{10}|\le\norm Z_1\).
Applying Eq.~\eqref{eq:blocktrace} to \(K_0\) and \((Q_0)_{\mathcal L}\), respectively, yields
\begin{equation}
 4\zeta\le\Tr A+\Tr B\le4-2\norm{t\diag(x,y)+N}_1.
\end{equation}
From the positive semidefiniteness of the compression of \(M\) on \(\mathcal O\oplus\operatorname{span}\{f\}\), and then using the upper bounds on the diagonal entries given by \(Q_0\succeq0\) at positions \(00\) and \(11\), we obtain
\begin{equation}
 |\Tr N|^2\le p\langle\Omega|U_0|\Omega\rangle\le2pt^2.
\end{equation}
Combining this with \(x,y\ge\sqrt{1-t^2}\), we further obtain
\begin{equation}
 \zeta\le1-t\sqrt{1-t^2}+t\sqrt{p/2},\qquad
 w\ge\frac{191}{100}t.
 \label{eq:linear-deficit}
\end{equation}
Among these, the second inequality can be proved by contradiction. If \(w<191t/100\), then Eq.~\eqref{eq:pbound} gives \(p<2t\), and hence
\(\sqrt{p/2}<\sqrt t<1/25\).
Using \(\sqrt{1-t^2}\ge1-t^2\) and denoting \(d=24/25-t^2\in(0,1)\), we then have
\begin{equation}
 w\ge(2d-d^2t)t\ge(48/25-t-2t^2)t>191t/100,
\end{equation}
contradicting the assumption.

\subsection{Schur complements and a positive definite comparison}
The proof is divided into three steps: first take the Schur complement with respect to the \(\mathcal S\) block, then estimate the \(\mathcal L\) block using a positive definite block-diagonal matrix, and finally prove that the scalar Schur complement corresponding to \(f\) is strictly positive. The second step does not require inverting \(K_0\), and therefore also applies when \(K_0\) is singular.

Write \(P_{\mathcal S,(a,\mathcal L,f)}=(b,W_S,q)\) and \(P_{a\mathcal L}=W_a\).
Decomposing \(U_0\) as a sum of rank-one positive semidefinite matrices gives \(\norm{U_0^\Gamma}\le\Tr U_0\). From \(M\succeq0\), we also have
\(\norm{M_{\mathcal O,\mathcal L}}_F^2\le\Theta\Tr(A+B)\le4\Theta\).
The partial transpose only changes the positions of these matrix elements and may take their complex conjugates, so
\begin{align}
 P_{\mathcal S}&\succeq d_tI,\qquad d_t=2t-\Theta>0,\nonumber\\
 \norm b&\le\Theta,\qquad \norm{W_S}_F^2+\norm{W_a}^2\le4\Theta.
 \label{eq:armbudget}
\end{align}
Using the weighted Cauchy--Schwarz inequality with weights \(1/24,1/24,11/12\) that sum to one, we obtain
\begin{equation}
 \begin{split}
 \norm{b\xi+W_S\ell+q\varphi}^2\le{}&24\Theta^2|\xi|^2
 +24\norm{W_S\ell}^2\\
 &+\tfrac{12}{11}\norm q^2|\varphi|^2.
 \end{split}
\end{equation}
After taking the Schur complement with respect to the \(\mathcal S\) block, we then use
\begin{equation}
 2\Re(\bar\xi W_a\ell)\ge-t|\xi|^2/10-10\norm{W_a\ell}^2/t
\end{equation}
to estimate the cross terms. In this way, the total amount subtracted from the \(\mathcal L\) block does not exceed
\begin{equation}
 \frac{24}{d_t}(W_S^\dagger W_S+W_a^\dagger W_a)
 \preceq\frac{50\Theta}{t}I.
\end{equation}
At the same time, \(P_{aa}\ge2-2t-t^2-\Theta\), and \(24\Theta^2/d_t\le t^2\).
Combining these estimates, the resulting Schur complement is no less than
\begin{equation}
 \begin{pmatrix}
 a_t&0&\sqrt2z\\0&\mathcal K&h'\\
 \sqrt2\bar z&(h')^\dagger&1-t^2+p-F_t
 \end{pmatrix},
 \label{eq:lowerblock}
\end{equation}
where
\begin{align}
 a_t&=2-\tfrac{21}{10}t-6t^2,\qquad
 \mathcal K=P_{\mathcal L\mathcal L}-\tfrac{50\Theta}{t}I,\nonumber\\
 F_t&=\tfrac{33}{40}t-\tfrac{\Theta}{4t}.
\end{align}
Here, the estimate for the last diagonal entry is guaranteed by Eq.~\eqref{eq:jointbudget}. Specifically,
\begin{align}
 d_tF_t-\tfrac{12}{11}\norm q^2
 &\ge t^2\left(\frac3{220}-\frac{12}{11}t\right)\nonumber\\
 &\quad+\Theta\left(\frac1{22}-\frac{33}{40}t\right)
       +\frac{\Theta^2}{4t}\ge0.
\end{align}
Since \(d_t>0\), we have \((12/11)\norm q^2/d_t\le F_t\).
Moreover, \(a_t>0\), and \(2/a_t\le1+53t/50\). The latter inequality can be verified by
\begin{equation}
 \frac{(1+53t/50)a_t-2}{t}
 =\frac1{50}-\frac{4113}{500}t-\frac{159}{25}t^2
 \ge\frac1{50}-9t>0.
\end{equation}

Next we estimate \(\mathcal K\). Let \(D=\diag(I+A,I+B)\) and \(D'=\diag(I+A,I+B^{\mathsf T})\).
From \(\norm Z\le1\) it follows that
\begin{equation}
 D-K_0=\begin{pmatrix}I&-Z\\-Z^\dagger&I\end{pmatrix}\succeq0.
\end{equation}
Let \(h=(h_A,h_B)^{\mathsf T}\). Adding \(\diag(D-K_0,0)\) to \(M_{(\mathcal L,f)}\) yields a positive semidefinite matrix whose upper-left block is \(D\). Since \(D\succ0\), taking the Schur complement with respect to this block gives
\begin{equation}
 (h')^\dagger(D')^{-1}h'=h^\dagger D^{-1}h\le p.
\end{equation}
Let \(e=50\Theta/t+3t\le203t<1/3\); then
\begin{align}
 \mathcal K&\succeq D'-eI\succeq(1-e)D'\succ0,\nonumber\\
 (h')^\dagger\mathcal K^{-1}h'&\le\frac p{1-e}.
 \label{eq:inverseclaim}
\end{align}
This estimate also holds for singular \(K_0\).

Finally, let \(\gamma=e/(1-e)\le3e/2<1/2\). In the matrix given by Eq.~\eqref{eq:lowerblock}, take the Schur complement with respect to \(\diag(a_t,\mathcal K)\), and denote the resulting scalar by \(\mathcal S_f\). Using Eqs.~\eqref{eq:pbound} and~\eqref{eq:linear-deficit}, we obtain
\begin{align}
 \mathcal S_f
 &\ge1-t^2-F_t-\gamma p-2\zeta^2/a_t\nonumber\\
 &\ge(1-\gamma)w-\tfrac{377}{200}t
       +\tfrac{\Theta}{4t}-(1+\gamma)t^2\nonumber\\
 &\ge\tfrac{t}{40}+\tfrac{\Theta}{4t}-2\gamma t-\tfrac32t^2\nonumber\\
 &\ge\frac{t}{40}-\frac{21}{2}t^2
       +\left(\frac1{4t}-150\right)\Theta
 \ge\frac{11}{1280}t>0.
 \label{eq:finalschur}
\end{align}
The last step uses \(1/(4t)-150\ge10\) and \(1/40-(21/2)t\ge11/1280\).
The matrix block eliminated at each Schur complement is positive definite, and the final scalar obtained is also strictly positive; therefore Lemma~\ref{lem:boundary} is proved. The explicit lower bound given by Eq.~\eqref{eq:finalschur} applies to this final scalar Schur complement.

\section{Dependence on the local dimension}
Theorem~\ref{thm:main} shows that all quantum states in bipartite three-level systems terminate before the same level. This is not in conflict with the counterexample given in Ref.~\cite{Brinster2026}. Theorem~8 of that paper constructs, for every \(m\ge3\), a quantum state on \(\mathbb C^{2^m}\otimes\mathbb C^{2^m}\) such that \(\chi_{m-2}<\chi_{m-1}\). Every state in this family terminates at level \(m-1\), but the termination level increases as the local dimension grows.

When \(m=4\), Eqs.~(C54)--(C56) of Ref.~\cite{Brinster2026} give
\begin{equation}
 \chi_3(\rho_4)=\frac83,\qquad\chi_2(\rho_4)\le\frac52,
\end{equation}
where the local dimensions on both sides are sixteen. Therefore, if one defines the smallest local dimension that can separate the second and third levels as
\begin{equation}
\begin{aligned}
 d_{\min}=\min\bigl\{d:\;&\exists\rho\text{ is a quantum state}\\
 &\text{on }\mathbb C^d\otimes\mathbb C^d \text{ with } \chi_3(\rho)>\chi_2(\rho)\bigr\}.
\end{aligned}
\end{equation}
then
\begin{equation}
                         4\le d_{\min}\le16.
\end{equation}
To establish the lower bound, it suffices to isometrically embed states with smaller local dimensions into the bipartite three-level space. Local isometric embedding leaves the values at each level unchanged. Indeed, embedding an original feasible chain in the original space into the larger space gives one inequality; conversely, compressing a feasible chain in the larger space onto the product subspace corresponding to the embedding gives the reverse inequality.

By Corollary~\ref{cor:cost}, the entanglement cost of any two-qutrit state is the logarithm of the optimal value of a fixed SDP. Thus there is no error due to truncation of the hierarchy, although numerical solutions of the SDP are still subject to optimization error. This result gives an upper bound on the termination level of the $\chi$ hierarchy. It does not determine whether the first level is sufficient for all two-qutrit states.

The proof uses the dimensional condition in two places. The finite-certificate part relies on positive semidefiniteness on the three-dimensional antisymmetric subspace, and this estimate uses the fact that every vector in this subspace has Schmidt rank at most two. At higher local dimensions, the antisymmetric subspace no longer has this property. The analytic part handles the case near \(t=0\) for vectors with three Schmidt coefficients. Therefore, the existing proof cannot yet extend the conclusion to all \(3\times d\) systems.

\section{Conclusion}
We prove that for any bipartite three-level quantum state, the semidefinite programming lower-bound hierarchy for the PPT entanglement cost terminates at the second level. Therefore, solving an SDP containing only three Hermitian matrix variables determines its zero-error asymptotic entanglement cost. We shorten the dual chain by one level using finitely many rational certificates away from the boundary and an analytic estimate near the boundary. The two arguments cover the entire parameter domain.

Several questions remain open. What is the smallest local dimension for which a strict separation occurs between the second and third levels of the lower-bound hierarchy? Does the second level of the upper hierarchy give the exact entanglement cost of every two-qutrit state? It would also be useful to find an analytic inequality that replaces the finite certificates used in the proof.

\noindent\textit{Note added.}---Approximately one month after posting the first version of this manuscript on arXiv, we became aware of subsequent work by Mele, Sen, and Lami~\cite{Mele2026}, who proved finite termination of the lower-bound hierarchy for the exact PPT entanglement cost in arbitrary finite dimensions. Their general bound gives termination at level \(\lfloor d_A d_B/2\rfloor\), hence at level four for two qutrits, whereas our earlier result establishes a lower uniform termination level of two for all two-qutrit states.

\begin{acknowledgments}
This work was supported by the Natural Science Foundation of Anhui Province under Grant No. 2508085ZD001, and Scientific Research Projects of Anhui Provincial Department of Education under Grant Nos. 2025AHGXZK20126 and 2025AHGXZK50054. The authors acknowledge the assistance of ChatGPT 6 Astra with the computational aspects of certificate construction and verification. The authors take full responsibility for the accuracy and integrity of this work.
\end{acknowledgments}

\appendix
\section{Exact verification of the certificates}
\label{app:verification}
The finite-certificate part uses the \(126\) groups of certificates introduced in Sec.~\ref{sec:finite}. The supplementary material~\cite{Supplement} provides all certificate data and the verification programs, which can be used to recompute every bound in the proof using rational arithmetic. The entire verification procedure does not rely on floating-point eigenvalue computations, nor does it use the output of an SDP solver.

Each Choi matrix is stored via an integer Gram factor. For the middle and upper regions, dividing the factor by \(10^8\); for the regions corresponding to the binary partition intervals, dividing the factor by \(10^{10}\), yields the rational matrices used in the proof. The positive semidefiniteness of the resulting Gram matrices is guaranteed exactly by this decomposition. The scalar certificates in Eq.~\eqref{eq:scalarcertificate} are also stored as Gram matrices in the same way, and the remaining scalar weights are verified one by one to be nonnegative. The program also recomputes the coefficient residuals, without presupposing that these rational data satisfy the target identities without error.

For each matrix variable, the program computes the values of the corresponding coefficient map on the \(45\) matrix units with \(i\le j\), and the remaining values are determined by Eq.~\eqref{eq:paired}. If the coefficient error corresponding to an off-diagonal matrix unit is \(A\), then its contribution to row \(r\) is estimated by the upper bound \(\sum_s(|A_{rs}|+|A_{sr}|)\). Thus, each pair of adjoint terms is included, and all off-diagonal entries in a complex Hermitian input are also handled. The routine verification requires \(126\times3\times45=17010\) computations in total. In addition, we directly compute the values on the remaining \(13608\) matrix units and check the nine row sums one by one against the results of the adjoint-pairing method. The residual bounds obtained by the two methods are exactly the same.

Subsequently, combining the scalar polynomial errors and the row-sum upper bounds for the off-diagonal entries according to Eq.~\eqref{eq:certificatemargin}, all resulting lower bounds are no less than the values listed in Table~\ref{tab:certificates}. The region coverage is verified separately: the program traverses all nodes of the partition tree for the middle region, confirms that every retained leaf node has a certificate, and checks whether every excluded leaf node satisfies the corresponding infeasibility inequality. Together with the six consecutive closed intervals and the upper cube, this covers the entire parameter domain handled by the finite certificates. Its lower endpoint \(1/640\) exactly matches the upper endpoint of the interval handled by Lemma~\ref{lem:boundary}.

The accompanying programs also verify the elementary rational inequalities in Sec.~\ref{sec:boundary}. They check partial transposition on all $81$ matrix units, the identities for the symmetric projector, and the index permutations used in the Frobenius norm estimates. If a transpose or a leaf of the partition tree is deliberately omitted, the corresponding check fails. These checks support the analytic argument near the boundary. The matrix inequalities and Schur complement steps are proved in the main text.


\begin{thebibliography}{99}
\bibitem{Hayden2001}
P. M. Hayden, M. Horodecki, and B. M. Terhal,
The asymptotic entanglement cost of preparing a quantum state,
\href{https://doi.org/10.1088/0305-4470/34/35/314}{J. Phys. A: Math. Gen. \textbf{34}, 6891 (2001)}.
\bibitem{Rains2001}
E. M. Rains, A semidefinite program for distillable entanglement,
\href{https://doi.org/10.1109/18.959270}{IEEE Trans. Inf. Theory \textbf{47}, 2921 (2001)}.
\bibitem{Audenaert2003}
K. Audenaert, M. B. Plenio, and J. Eisert,
Entanglement cost under positive-partial-transpose-preserving operations,
\href{https://doi.org/10.1103/PhysRevLett.90.027901}{Phys. Rev. Lett. \textbf{90}, 027901 (2003)}.
\bibitem{Wang2020}
X. Wang and M. M. Wilde,
Cost of quantum entanglement simplified,
\href{https://doi.org/10.1103/PhysRevLett.125.040502}{Phys. Rev. Lett. \textbf{125}, 040502 (2020)}.
\bibitem{Wang2023}
X. Wang and M. M. Wilde,
Exact entanglement cost of quantum states and channels under positive-partial-transpose-preserving operations,
\href{https://doi.org/10.1103/PhysRevA.107.012429}{Phys. Rev. A \textbf{107}, 012429 (2023)}.
\bibitem{Wang2024Errata}
X. Wang and M. M. Wilde,
Errata for ``Cost of quantum entanglement simplified'' and ``Exact entanglement cost of quantum states and channels under PPT-preserving operations,''
Zenodo, version 2 (2024), \href{https://doi.org/10.5281/zenodo.11061607}{doi:10.5281/zenodo.11061607}.
\bibitem{Lami2025}
L. Lami, F. A. Mele, and B. Regula,
Computable entanglement cost under positive partial transpose operations,
\href{https://doi.org/10.1103/PhysRevLett.134.090202}{Phys. Rev. Lett. \textbf{134}, 090202 (2025)}.
\bibitem{Brinster2026}
R. Brinster, T. V. Kondra, H. Kampermann, D. Bru{\ss}, and N. Wyderka,
Multinegativity and single-letter formulas for asymptotic entanglement,
\href{https://arxiv.org/abs/2609.20698}{arXiv:2609.20698} (2026).
\bibitem{Vandenberghe1996}
L. Vandenberghe and S. Boyd,
Semidefinite programming,
\href{https://doi.org/10.1137/1038003}{SIAM Rev. \textbf{38}, 49 (1996)}.
\bibitem{Choi1975}
M.-D. Choi,
Completely positive linear maps on complex matrices,
\href{https://doi.org/10.1016/0024-3795(75)90075-0}{Linear Algebra Appl. \textbf{10}, 285 (1975)}.
\bibitem{Supplement}
See Supplemental Material at https://github.com/wei-song527/qutrit-ppt-cost.
\bibitem{Mele2026}
F. A. Mele, S. Sen, and L. Lami,
Cost of entanglement simplified: a single-letter formula (after all),
\href{https://arxiv.org/abs/2611.41528v1}{arXiv:2611.41528v1} (2026).
\end{thebibliography}
\end{document}